\documentclass[11pt]{article}
\usepackage{amsmath}
\usepackage{graphicx}
\usepackage{amssymb}
\usepackage{amsfonts}
\usepackage{latexsym}
\usepackage[pdftex,pdfmenubar=true,bookmarks=true,pdftoolbar=true]{hyperref}
\hypersetup{colorlinks,citecolor=green,filecolor=magenta,linkcolor=red,urlcolor=cyan,pdftex}
\def\be{\begin{equation}} \def\ee{\end{equation}}
\def\ba{\begin{eqnarray}} \def\ea{\end{eqnarray}} \def\part{\partial}

\begin{document}

%%%%%%%%%%%%%%%%%%%%%%%%%%%%%%%%%%%%%%%%%%%%%%%%%%%%%%%%%%%%%%%%%%%%%%%%
\begin{center}
\begin{flushright}\begin{small}    UFES 2012
\end{small} \end{flushright} \vspace{1.5cm}
\huge{Thermodynamics of a class of non-asymptotically flat black holes in Einstein-Maxwell-Dilaton theory} 
\end{center}

\begin{center}
{\small \bf Manuel E. Rodrigues $^{(a,b,c)}$}\footnote{E-mail
address: esialg@gmail.com}\ and
{\small \bf Glauber Tadaiesky Marques $^{(a)}$}\footnote{E-mail address:
gtadaiesky@hotmail.com} \vskip 4mm

(a) \ Universidade Federal Rural da Amaz\^{o}nia-Brazil\\
ICIBE - LASIC\\
Av. Presidente Tancredo Neves 2501\\ CEP66077-901 -
Bel\'em/PA, Brazil
\vskip 2mm (b) \ Universidade Federal do Esp\'{\i}rito Santo \\
Centro de Ci\^{e}ncias
Exatas - Departamento de F\'{\i}sica\\
Av. Fernando Ferrari s/n - Campus de Goiabeiras\\ CEP29075-910 -
Vit\'{o}ria/ES, Brazil \\
\vskip 2mm(c) \ Faculdade de F\'{\i}sica, Universidade Federal do Par\'{a}, 66075-110, Bel\'em, Par\'{a}, Brazil
\vskip 2mm

\end{center}

%%%%%%%%%%%%%%%%%%%%%%%%%%%%%%%%%%%%%%%%%%%%%%%%%%%%%%%%%%%%%%%%%%%%%%%%%%%%%%%%%%%%%%%%%%%%%%%%%%%%%%%%%%%%%%%%%%%%%%%%%%%%%%%%%%%%%%%%%%%%%%%%%%%%%%%
\begin{center}
                                       Abstract
\end{center}
\hspace{0,4cm} 
We analyse in detail the thermodynamics in the canonical and grand canonical ensembles of a class of non-asymptotically flat black holes of the Einstein-(anti) Maxwell-(anti) Dilaton theory in 4D with spherical symmetry. We present the first law of thermodynamics, the thermodynamic analysis of the system through the geometrothermodynamics methods, Weinhold, Ruppeiner, Liu-Lu-Luo-Shao and the most common, that made by the specific heat. The geometric methods show a curvature scalar identically zero, which is incompatible with the results of the analysis made by the non null specific heat, which shows that the system is thermodynamically interacting, does not possess extreme case nor phase transition. We also analyse the local and global stability of the thermodynamic system,  and obtain a local and global stability for the normal case for $0<\gamma<1$ and for other values of $\gamma$, an unstable system. The solution where $\gamma=0$ separates the class of locally and globally stable solutions from the unstable ones.

\vspace{0,5cm}

Pacs numbers: 04.70.-s; 04.20.Jb; 04.70.Dy.

%%%%%%%%%%%%%%%%%%%%%%%%%%%%%%%%%%%%%%%%%%%%%%%%%%%%%%%%%%%%%%%%%%%%%%%%%%%%%%%%%%%%%%%%%%%%%%%%%%%%%%%%%%%%%%%%%%%%%%%%%%%%%%%%%%%%%%%%%%%%%%%%%%%%%%%

\section{Introduction}

\hspace{0,6cm} The discovery that a black hole can radiate a temperature characteristic of a blackbody spectrum, was made by Hawking. This opened the possibility of constructing a semi-classical thermodynamics for systems of solutions of black holes arising from general relativity and its recent modifications.
\par
An interesting class of black holes is that which comes from the minimal coupling in 4 dimensions, from the Einstein-Hilbert action with a scalar field and Maxwell field, commonly called the Einstein-Maxwell-Dilaton (EMD) theory. This theory, imposing spherical symmetry and a static metric, can provide non-asymptotically flat (NAF) solutions. These solutions have several characteristics that have not yet been studied with great detail. Moreover, we also observe that their thermodynamics is not completely established. This class of solutions was first obtained in \cite{horne}. Several generalizations have been made, as the case with rotation \cite{mitra}, d-dimensional \cite{kiem}, multicenter solutions \cite{gerard1}, black branes on the linear dilaton background \cite{gerard2} and phantom black holes with degenerate horizon \cite{gerard3}. Here, we will widely study the thermodynamics of the class of solutions $(2.27)$ and $(2.28)$ of \cite{glauber}.
\par
Several other solutions of  non asymptotically flat or non asymptotically (A)dS black holes have were studied in detail in \cite{mignemi}-\cite{yazadjiev}. These solutions attract much attention due to the possibility of further elucidate the understanding of the AdS/CFT correspondence, regarding the linear dilaton solutions,  arising in a near horizon limit \cite{gerard5}, which can lodge holography \cite{aharony}. One of the generalizations of the usual black hole solutions is for a non trivial topology, as in \cite{sheykhi5}, this enables us different thermodynamic properties of the usual. The Hawking-Page phase transition \cite{page} is an important thermodynamics property for AdS-type black holes, which does not seem to occur to these generalizations. Here, we will also do this analysis as a means of comparison with other results previously established.
\par
There are some methods of analysis in black hole thermodynamics theory dubbed as geometrical, because they make use of differential geometry to determine thermodynamic properties such as: points at which black holes become extremal or they pass through a phase transition and thermodynamic stability  of the system. One of the first methods was proposed by Rao \cite{rao}, subsequently developed by other authors \cite{amari}. Later, the work of Weinhold \cite{weinhold} and Ruppeiner \cite{ruppeiner} were frequently used for the study of the black hole thermodynamics. For analysing the thermodynamics of the neither asymptotically flat nor (A)dS class of black holes, we use four methods called geometrics, the Weinhold \cite{weinhold}, Ruppeiner \cite{ruppeiner}, Geometrothermodynamics (GTD) \cite{quevedo} and that of Liu-Lu-Luo-Shao \cite{liu} methods. Regarding the latter two methods, they are in a good concordance when different thermodynamic potentials are chosen, like for example the mass and entropy representations, this in virtue of the invariance of the formalism by Legendre transformations. Finally, we have mention that the results are independent of the particular thermodynamic ensemble considered.
\par
However, we also have to point out that the GTD method can contain some inconsistencies when compared with the more usual analysis done by the specific heat. Recently it has shown that for the cases Reissner-Nordstrom-AdS and (phantom case) anti-Reissner-Nordstrom-AdS black holes \cite{deborah}, the GTD method does not reproduce the results obtained by the specific heat method. 
\par
For analysing the thermodynamics of the neither asymptotically flat nor (A)dS class of black holes, we use four methods called geometrics, as well as the more usual method that analyses the specific heat of the black hole.
\par
The paper is organized as follows. In Section \ref{sec2}, we re-obtain a class of non asymptotically flat black hole solutions of the EMD theory and its thermodynamics variables. We establish the first law of black holes thermodynamics and calculate its specific heat. In Section \ref{sec3}, we divide the analysis of the thermodynamic system through the methods of specific heat, subsection \ref{subsec3.1}, geometrotermodynamics, subsection \ref{subsec3.2},  Weinhold, subsection \ref{subsec3.3},  Ruppeiner, subsection \ref{subsec3.4},  Liu-Lu-Luo-shao, subsection \ref{subsec3.5} and the local and global stabilities in the subsection \ref{subsec3.6}. The conclusion and perspectives are presented in  Section \ref{sec4}.  

%%%%%%%%%%%%%%%%%%%%%%%%%%%%%%%%%%%%%%%%%%%%%%%%%%%%%%%%%%%%%%%%%%%%%%%%%%%%%%%%%%%%%%%%%%%%%%%%%%%%%%%%%%%%%%%%%%%%%%%%%%%%%%%%%%%%%%%%%%%%%%%%%%%%

\section{ The field equations and the black hole solutions}\label{sec2}

\hspace{0,5 cm} For understanding very well the origin of the parameters  and structure of the solutions discussed here, we will construct  the technique of obtaining a classes of solutions previously found, as shown in \cite{gerard3}.

The action of EMD theory is given by: 
\be
S=\int dx^{4}\sqrt{-g}\left[  \mathcal{R}-2\,\eta_{1}
g^{\mu\nu}\nabla_{\mu}\varphi\nabla_{\nu }\varphi+\eta_{2} \,e^{
2\lambda\varphi}F^{\mu\nu}F_{\mu\nu}\right]  \label{action1}\; ,
\end{equation}
where the first term is the usual Einstein-Hilbert gravitational term, while the second and the third are respectively a kinetic term of the scalar field (dilaton or phantom) and a coupling term between the scalar and the Max\-well fields, with a coupling constant $\lambda$ that we assume to be real. The coupling constant $\eta_1$ can take either the value $\eta_{1}=1$ ({\it dilaton}) or $\eta_{1}=-1$ ({\it anti-dilaton}). The Maxwell-gravity coupling constant $\eta_{2}$ can take either the value $\eta_{2}=1$ ({\it Maxwell}) or $\eta_{2}=-1$ ({\it anti-Maxwell}). This action leads to the following field equations:
\begin{eqnarray}
\nabla_\mu\left[  e^{2\lambda\varphi}
F^{\mu\alpha}\right]&=&0\label{fe1}\; ,\\
\Box\varphi &=& - \frac{1}{2}\eta_{1}\eta_{2}\lambda e^{2\lambda\varphi}F^2
\; ,\label{fe2}\\
R_{\mu\nu}&=&2\eta_{1}\nabla_{\mu}\varphi\nabla_{\nu}\varphi+2\eta_{2}\,
e^{2\lambda\varphi}\left(\frac {1}{4}g_{\mu\nu}F^{2}
-F_{\mu}^{\;\;\sigma}F_{\nu\sigma}\right)\label{fe3}\, .
\end{eqnarray}

From now, we will make use of the same procedure as in \cite{gerard3}. Let us write the static and spherically symmetric line element as
\begin{equation}
dS^{2}=e^{2\gamma(u)}dt^{2}-e^{2\alpha(u)}du^{2}-e^{2\beta(u)}d\Omega^{2}
\label{metric}\; .
\end{equation}

The metric function $\alpha$ can be changed  according to the redefinition of the radial coordinate $u$. Then, we consider the harmonic
coordinate condition
\be 
\alpha = 2\beta + \gamma \,.
\ee

We will also assume that the Maxwell field is purely electric (the purely
magnetic case may be obtained by electromagnetic duality transformation $\varphi\rightarrow -\varphi$, $F \to
e^{-2\lambda\varphi}*F$). Integrating (\ref{fe1}), we obtain
\begin{equation}
F ^{10}(u)=q e^ {-2(\lambda\varphi+2\beta+\gamma)} \qquad
(F ^{2}=-2q^{2}e^{-4\beta-4\lambda\varphi}) \label{s1}\; ,
\end{equation}
with $q$ a real integration constant. Substituting  (\ref{s1}) into the equations of motion, we
obtain the equations of second order 
\begin{eqnarray}
\varphi''  &=& -\eta_{1}\eta_{2}\lambda q ^{2}e^{2\omega} \label{e2bis}\; ,\\
\gamma'' &=&\eta_{2}q ^{2}e^{2\omega} \label{e1bis}\; ,\\
\beta'' &=&e^{2J}-\eta_{2}q ^{2}e^{2\omega} \label{e3bis}\; ,
\end{eqnarray}
with
\be
\omega=\gamma-\lambda\varphi\,, \quad J=\gamma+\beta \,,
\ee
and the constraint equation
\be\label{cons}
\beta^{'2} + 2\beta'\gamma' - \eta_1\varphi'^2 = e^{2J} - \eta_2q^2
e^{2\omega}\,.
\ee

By taking linear combinations of the equations (\ref{e2bis})-(\ref{e3bis}),
this system can be integrated and one gets
\begin{eqnarray}
\varphi (u)&=&-\eta_{1}\lambda\gamma (u) +\varphi_{1}u+\varphi_{0}\; ,
\label{e1tertio}\\
\omega'^2 -Qe^{2\omega}&=& a^2\label{e2tertio}\; ,\\
J'^{2}-e^{2J}&=&b^{2}\label{e3tertio}\; ,
\end{eqnarray}
where
\be
\lambda_{\pm}=(1\pm\eta_{1}\lambda^{2})\,, \quad Q=\eta_{2}\lambda_{+}q^2\,,
\ee
and the integration constants $\varphi_{0},\varphi_{1}\in\mathbb{R}$,
$a,b\in\mathbb{C}$.

The general solution of (\ref{e2tertio}) is:
\begin{eqnarray}\label{om}
\omega(u)=\left\{\begin{array}{lr}
-\ln \left| \sqrt{\left| Q\right|} a^{-1}\cosh [a(u-u_{0})]\right| \quad
(a\in \mathbb{R}^{+}\; , \;Q\in\mathbb{R}^{-})\; , \\
a(u-u_{0}) \qquad (a\in\mathbb{R}^{+}\;, \; Q=0 )\; ,\\
-\ln \left| \sqrt{Q} a^{-1}\sinh [a(u-u_{0})]\right| \quad (a\in \mathbb{R}^+
\;,  \;Q\in\mathbb{R}^{+})\; ,\\
-\ln \left| \sqrt{Q} (u-u_{0})\right| \quad (a=0,\; Q\in\mathbb{R}^{+})\;,\\
-\ln \left| \sqrt{Q} \bar{a}^{-1}\sin [\bar{a}(u-u_{0})]\right| \quad (a
=i\bar{a},\; \bar{a},Q\in\mathbb{R}^{+})
\end{array}\right.
\end{eqnarray}
(with $u_0$ a real constant). The general solution of (\ref{e3tertio}) reads:
\begin{equation}\label{J}
J(u)=\left\{\begin{array}{lr}
-\ln \left|  b^{-1}\sinh [b(u-u_{1})]\right| \quad (b\in\mathbb{R}^{+})\;,\\
-\ln \left| u-u_{1}\right| \quad (b=0)\; ,\\
-\ln \left| \bar{b}^{-1}\sin [\bar{b}(u-u_{1})]\right| \quad (b=i\bar{b} ;\;
\bar{b}\in\mathbb{R}^{+})
\end{array}\right.
\end{equation}
($u_1$ real constant). In this way, we have for $\lambda_+\neq 0$, the static and spherically  general solution of
the theory given by the action (\ref{action1}):
\begin{eqnarray}
\label{gs} \left\{\begin{array}{lr}
dS^{2}=e^{2\gamma}dt^{2}-e^{2\alpha}du^{2}-e^{2\beta}d\Omega^{2}\; ,\\
\alpha (u)=2J(u)-\gamma (u)\; ,\\
\beta (u)=J(u)-\gamma (u)\; , \\
\gamma (u)=\lambda _{+}^{-1}(\omega(u)+\lambda\varphi_{1}u+\lambda\varphi_0)
\; ,\\
\varphi (u)=\lambda _{+}^{-1}(-\eta_{1}\lambda\omega(u)+\varphi _{1}u+
\varphi _{0})\; ,\\
F=-q\; e^{2\omega (u)}du\wedge dt\;,
\end{array}\right.
\end{eqnarray}
where, by fixing the spacial infinity at $u_1=0$, we have six integrations constants ($ q, a, b, u_{0}, \varphi _{0}, \varphi_1$), which obey the following  constraint equation (\ref{cons}):
\begin{equation}
\lambda _{+}b^{2}=a^{2}+\eta _{1}\varphi
_{1}^{2}\label{constraint}\; .
\end{equation}

Now, we will choose one particular class of the solution of that obtained in \cite{gerard3}. According to \cite{gerard3}, if we fix the solutions of $\omega(u)$ and $J(u)$ to be $\sinh$ functions, with the non-degenerated horizon ($m=n=1$) and $u_0=0$, and making the change of radial coordinate  $e^{2bu}=f_+$, where $f_+=1-(r_+/r)$, we obtain the following solution  
\begin{eqnarray}
dS^{2}=\frac{r^{\gamma}(r-r_+)}{r_{0}^{1+\gamma}}dt^{2}-\frac{r_{0}^{1+\gamma}%
}{r^{\gamma}(r-r_+)}dr^{2}-r_{0}^{1+\gamma}r^{1-\gamma}d\Omega^{2}\, , \label{6}\\
F=\mp\sqrt{\frac{1+\gamma}{2\eta_2}}\frac{1}{r_0}dr\wedge dt \;\; ,\;\; e^{2\lambda\varphi}=\left(\frac{r}{r_0}\right)^{1-\gamma}\; ,\label{7}
\end{eqnarray}
where the parameter $\gamma=(1-\eta_1\lambda^2)/(1+\eta_1\lambda^2)$. This is an exact solution of a spherically symmetric, non asymptotically flat (NAF), static and electrically charged black hole, with event horizon $r_+$ ($r_+\geq 0$). The parameters $r_+$ and $r_0$ ($r_0\geq 0$) are related to the physical mass\footnote{We use the quasi-local mass defined in \cite{booth,hawking2}, because it is a non asymptotically flat spacetime.} and the electric charge by:
\begin{align}
M &  =\frac{\left(  1-\gamma\right)  }{4}r_+\label{8}\; ,\\
q &  =\pm r_{0}\sqrt{\frac{\left(  1+\gamma\right)  }{2\eta_2}}\, .\label{9}
\end{align}

Here, the parameter $r_0$ is related to the electric charge. From here, we must take the sign ($+$) in  (\ref{9}), which implies in taking the sign  ($-$) in (\ref{7}), for the Maxwell field. This does not mean that we do not treat the solution with the sign ($-$) from $q$ ($q<0$), but rather, we do not write explicitly the two possibilities, with $q\in\mathbb{R}$. However, the calculations are valid for the two cases, without any loss of generality. We also notice that the phantom cases arising from the choice of the function $\omega(u)$ in (\ref{om}), being $\sinh$, and we impose $\eta_2(1+\eta_1\lambda^2)>0$. Then, the phantom cases are reduced to $\eta_1=-\eta_2=-1$ with $1<\gamma<+\infty$ and $\eta_1=\eta_2=-1$ for $-\infty <\gamma<-1$. But, since we need a positive mass for (\ref{8}), we just have the unique phantom case $\eta_1=\eta_2=-1$ for $-\infty <\gamma<-1$. Considering the normal and phantom cases, one gets the following interval $\gamma\in(-\infty,+1)$.
\par
Now, we are interested in the geometrical analysis representing semi-classical gravitational effects of the black hole solutions mentioned before. By semi-classical we mean quantize the called matter fields, while leaving classical the background gravitational field. Therefore we will work with the semi-classic thermodynamics of black holes, studied first by Hawking \cite{hawking1}, and further developed by many other authors \cite{davies2}.

There are several techniques to derive the Hawking temperature law. For example we can mention the Bogoliubov coefficients \cite{ford} and the energy-momentum tensor methods \cite{davies,davies2}, by the euclidianization of the metric \cite{hawking2}, the transmission and reflection coefficients \cite{gerard4,kanti}, the analysis of the anomaly term \cite{robinson}, and by the black hole superficial gravity \cite{jacobson}. Since all these methods have been proved to be equivalents \cite{glauber}, then we opt, without loss of generality, to calculate the Hawking temperature by the superficial gravity method.

The surface gravity of a black hole is given by \cite{wald}:
\begin{equation}
\kappa =\left[  \frac{g_{00}^{\prime}}{2\sqrt{-g_{00}g_{11}}}\right]_{r\;=\;r_{+}}\, ,\label{10}
\end{equation}
where $r_{+}$ is the event horizon radius, and the Hawking temperature is related with the surface  gravity through the relationship \cite{jacobson,hawking1}
\begin{equation}
T=\frac{\kappa}{2\pi}\, .\label{11}
\end{equation}

Then, for the NAF black hole case (\ref{6}), we get the surface gravity (\ref{10}) as 
\begin{equation}
 \kappa =\frac{r_{+}^{\gamma}}{2r_{0}^{1+\gamma}}\,  ,\label{12}
\end{equation}
and the Hawking temperature (\ref{11}), in this case is:
\begin{equation}
T=\frac{r_{+}^{\gamma}}{4\pi r_{0}^{1+\gamma}}\, .\label{13}
\end{equation}

We define the entropy of the black hole as a quarter of the horizon area as 
\begin{eqnarray}
S=\frac{1}{4}A=\frac{1}{4}\int \sqrt{g_{22}g_{33}}d\theta d\phi\Big|_{r=r_+}=\pi r_{0}^{1+\gamma}r_{+}^{1-\gamma}\,.\label{s}
\end{eqnarray}

From (\ref{7}), we can calculate the electric scalar potential on the horizon. But in this case, we can not directly integrate the component $F_{10}$ with respect to $r$, because the spacetime is NAF and provides an infinite contribution when $r\rightarrow +\infty$. This problem is known in literature \cite{gerard1}, and also happens to black holes in $(1+2)$-dimensions \cite{gerard6}. We can solve this problem by proposing an electric potential able to provide us the Maxwell field in (\ref{7}) and that satisfies the first law of thermodynamics of black holes. We then define the following electric potential on the horizon
\begin{eqnarray}
A_{0}=-\sqrt{\frac{1+\gamma}{2\eta_2}}\frac{r_+}{r_0}+a_0,\label{14}
\end{eqnarray}
where $a_0$ is a constant which fixes the potential gauge, from which we derive directly the Maxwell field in (\ref{7}). We will check the first law of thermodynamics for the solution (\ref{6}). Taking the differential of the mass (\ref{8}), of the electric charge  (\ref{9}) and of the entropy (\ref{s}), we obtain 
\begin{eqnarray}
dM=\frac{(1-\gamma)}{4}dr_+\,,\,dq=\sqrt{\frac{1+\gamma}{2\eta_2}}dr_0\label{dmass}\,,\\
dS=\pi (1+\gamma)r_{0}^{\gamma}r_{+}^{1-\gamma}dr_0+\pi (1-\gamma)r_{0}^{1+\gamma}r_{+}^{-\gamma}dr_{+}\label{ds}\,.
\end{eqnarray}

In order to to satisfy the first law of black hole thermodynamics,  
\begin{eqnarray}
dM=TdS+\eta_2 A_{0}dq\,,\label{pl}
\end{eqnarray}
it is necessary to fix the gauge with 
\begin{equation}
a_0=\frac{r_+}{2r_0}\sqrt{\frac{1+\gamma}{2\eta_2}}\;,
\end{equation}
and then, the electric potential (\ref{14})  is written as 
\begin{equation}
A_0=-\frac{r_+}{2r_0}\sqrt{\frac{1+\gamma}{2\eta_2}}\;.\label{15}
\end{equation}
Note that we introduced a compensatory $\eta_2$ in (\ref{pl}), due to the contribution of the negative energy density, in the phantom case, of the field of spin $1$, $F_{\mu\nu}$, which provides a work with inverted sign in the first law. For studying the thermodynamics of a geometric form, it is useful to write the temperature, entropy and the electric potential in functions of the mass and the electric charge. To do this, from (\ref{8}) and (\ref{9}), we get
\begin{eqnarray}
r_{0}=q\sqrt{\frac{2\eta_2}{1+\gamma}}\,,\, r_{+}=\frac{4M}{1-\gamma}\,,
\end{eqnarray}
which, from  (\ref{13}), (\ref{s}) and (\ref{15}) provides the temperature, the entropy and the electric potential in functions of the mass and the electric charge
\begin{eqnarray}
T&=&T_1M^{\gamma}q^{-(1+\gamma)}\,,\,T_1=\frac{2^{\frac{3\gamma-5}{2}}}{\pi}\frac{[\eta_2(1+\gamma)]^{\frac{1+\gamma}{2}}}{(1-\gamma)^{\gamma}}\,,\label{t1}\\
S&=&S_1 M^{1-\gamma}q^{1+\gamma}\,,\,S_1=\frac{\pi 2^{\frac{5-3\gamma}{2}}}{[\eta_2(1+\gamma)]^{\frac{1+\gamma}{2}}}(1-\gamma)^{\gamma-1}\,,\label{s1}\\
A_0&=&-\bar{A}_0\frac{M}{q}\,,\,\bar{A}_0=\eta_2\left(\frac{1+\gamma}{1-\gamma}\right)\label{a1}\,.
\end{eqnarray}
Here, we mention that the temperature and the entropy are always positive, since  we had taken the case  $q>0$, and the electric potential is always negative.

One can invert (\ref{s1}) for writing the mass in terms of the entropy and the electric charge 
\begin{eqnarray}
M(S,q)=q^{-\frac{1+\gamma}{1-\gamma}}\left(\frac{S}{S_1}\right)^{\frac{1}{1-\gamma}}\label{m1}\;.
\end{eqnarray}
Then, one can express the specific heat as 
\begin{equation}
C_{q}=\left(\frac{\partial M}{\partial T}\right)_{q}=\left(\frac{\partial M}{\partial S}\right)_{q}\Big/\left(\frac{\partial^2 M}{\partial S^2}\right)_{q}=\frac{(1-\gamma)}{\gamma}S\label{cqm}\;,
\end{equation} 
or equivalently by 
\begin{equation}
C_{q}=-\left(\frac{\partial S}{\partial M}\right)_{q}^2\Big/\left(\frac{\partial^2 S}{\partial M^2}\right)_{q}
=\frac{(1-\gamma)}{\gamma}S_{1}q^{1+\gamma}M^{1-\gamma}\label{cqs}\;.
\end{equation} 

Now,we have in hand all the ingredient for studying the thermodynamics of this specific system. In the next section we will analyse the specific heat and use four geometric methods for studying the thermodynamics of this class of NAF black holes. 
%%%%%%%%%%%%%%%%%%%%%%%%%%%%%%%%%%%%%%%%%%%%%%%%%%%%%%%%%%%%%%%%%%%%%%%%%%%%%%%%%%%%%%%%%%%%%%%%%%%%%%%%%%%%%%%%%%%%%%%%%%%%%%%%%%%%%%%%%%%%%%%%%%%%%%%%%%%%%%%%%%%%%%%%%%%%%%%%%%%%%%%%%%%%%%%%%%%%%%%%%%%%%%%%%%%%%%%%%%%%%%%%%%%%%%%%%%%%%%%%
\section{ Thermodynamics of NAF black holes}\label{sec3}
%%%%%%%%%%%%%%%%%%%%%%%%%%%%%%%%

\subsection{The usual thermodynamics analysis}\label{subsec3.1}
%%%%%%%%%%%%%%%%%%%%%%%%%%%%%%%%%%5
\par
Let us start our study of the thermodynamics of the class of NAF black hole solutions of EMD theory, using the analysis of the specific heat. The specific heat calculated by the use of the mass is directly proportional to the entropy, as shown in (\ref{cqm}). Then,  as we have a single horizon $r_+$, there are no extreme case nor phase transition. So, the geometric methods that present acceptable results must reproduce the result of the specific heat. In general, the thermodynamics system presents thermodynamic interaction, since both the specific heat and the temperature of the black hole are non null, but do not possess the extreme case, nor phase transition.
\par
In the next subsections, we will study the system via thermodynamic methods called geometrics. These methods consist in determining geometry, through a metric for the thermodynamic space of equilibrium states, resulting in a curvature scalar of this metric. This scalar can determine whether the system possesses phase transition and sometimes if there is an extreme case or thermodynamic interaction. We will make the calculations by using a mathematical software.
\par
In subsection \ref{subsec3.6} we will study the local stability by the specific heat studied here. In the next subsection we analyse thermodynamic system by the Geometrothermodynamics method.

%%%%%%%%%%%%%%%%%%%%%%%%%%%%%%%

\subsection{ The geometrothermodynamics method}\label{subsec3.2}
%%%%%%%%%%%%%%%%%%%%%%%%%%%%%%%
\hspace{0,6cm}The geometrothermodynamics (GTD) make use of differential geometry as a tool to represent the thermodynamics of physical systems. Let us consider the $(2n+1)$-dimensional space $\mathbb{T}$, which coordinates are represented by the thermodynamic potential $\Phi$, the extensive variable $E^{a}$ and the intensive variables $I^{a}$, where $a=1,...,n$. If the space $\mathbb{T}$ has a non degenerate metric $G_{AB}(Z^{C})$, where $Z^{C}=\{ \Phi , E^{a} , I^{a}\}$, and the so called Gibbs 1-form  $\Theta=d\Phi -\delta_{ab}I^{a}dE^{b}$, with $\delta_{ab}$ the delta Kronecker; then the structure $\left(\mathbb{T} , \Theta, G\right)$ is said to be a contact riemannian manifold if $\Theta\wedge \left(d\Theta\right)^{n}\neq 0$ is satisfied \cite{hermann}. The space $\mathbb{T}$ is known as the thermodynamic phase space. We can define a $n$-dimensional subspace $\mathbb{E}\subset \mathbb{T}$, with extensive coordinates $E^{a}$, by the map $\varphi :\mathbb{E}\rightarrow\mathbb{T}$, with $\Phi\equiv \Phi (E^{a})$, such that $\varphi^{*}(\Theta)\equiv 0$. We call the space $\mathbb{E}$ the thermodynamic space of the equilibrium states.
\par
We define the metric for the thermodynamic space of the equilibrium states $\mathbb{E}$  as \cite{zui}
\begin{align}
dl^{2}_{G(\Phi)} & =\left( E^{c}\frac{\partial\Phi}{\partial E^{c}}\right) \left( \eta_{ad}\delta^{di}\frac{\partial^{2}\Phi}{\partial E^{i}E^{b}}\right)dE^{a}dE^{b}\;,\label{me}
\end{align}
which is invariant under the Legendre transformations. The main interpretation of the thermodynamic system, made by this method is realised by calculating the curvature scalar of the metric of thermodynamic space of the equilibrium states (\ref{me}). With the curvature scalar we can identify whether there exists thermodynamic interaction in the system for non null scalar; if there is extreme black hole, when the scale and zero at a specific point, and finally if the system undergoes phase transitions.
\par
For the thermodynamic potential being the entropy (\ref{s1}), the metric (\ref{me}) has to be given by 
\begin{eqnarray}
dl^{2}_{G(S)}&=&\hspace{-0,3cm}\left(M\frac{\partial S}{\partial M}+q\frac{\partial S}{\partial q}\right)\left(-\frac{\partial^2 S}{\partial M^2}dM^2+\frac{\partial^2 S}{\partial q^2}dq^2\right)\\
&=&\hspace{-0,3cm}2S_{1}^{2}\gamma (1-\gamma)M^{-2\gamma}q^{2(1+\gamma)}dM^2+2S_{1}^{2}\gamma (1+\gamma)M^{2(1-\gamma)}q^{2\gamma}dq^{2}\label{mes}\;.
\end{eqnarray}
The curvature scalar of this metric is identically zero, so there is no thermodynamic interaction according to the interpretation of this method \cite{quevedo}. This means that the black hole could not radiate, being compared to an extreme black hole. This result is not in accordance with that of the specific heat, showing some inconsistency here in this method, for this specific case.
\par
Making the calculus for the metric of  $\mathbb{E}$, using  the mass  (\ref{m1}) as thermodynamic potential, on gets
\begin{eqnarray}
dl^{2}_{G(M)}&=&\frac{\gamma^2 q^{-2\frac{(1+\gamma}{(1-\gamma)}}}{S^2(1-\gamma)^3}\left(\frac{S}{S_1}\right)^{\frac{2}{1-\gamma}}dS^2-\frac{2\gamma (1+\gamma)}{(1-\gamma)^3}q^{-\frac{4}{1-\gamma}}\times\nonumber\\
&&\times\left(\frac{S}{S_1}\right)^{\frac{2}{1-\gamma}}dq^{2}\label{mem}\;.
\end{eqnarray}
The curvature scalar of this metric is also identically zero. Once again the incompatibility with the analysis of GTD with that of specific heat has been obtained. This was first shown in pathological solutions of phantom black holes \cite{zui}, next in  (anti) Reissner-Nordstrom-AdS black holes \cite{deborah}, and also shown here.

%%%%%%%%%%%%%%%%%%%%%%%%%%%%%%%

\subsection{ The Weinhold method}\label{subsec3.3}
%%%%%%%%%%%%%%%%%%%%%%%%%%%%%%%
One of the first to formulate a geometric analysis for a thermodynamic system was Weinhold. His method, known as the Weinhold method \cite{weinhold}, is to define a metric of the thermodynamic space of equilibrium states, through the mass as thermodynamic potential. This metric is used for the calculation of the curvature scalar $R_W$, which determine whether the system possesses phase transitions. The Weinhold metric is given by
\begin{eqnarray}
dl^{2}_{W(M)}&=&\frac{\partial^2 M}{\partial S^2}dS^2+2\frac{\partial^2 M}{\partial S\partial q}dSdq+\frac{\partial^2 M}{\partial q^2}dq^2\nonumber\\
&=&\frac{\gamma q^{-\frac{1+\gamma}{1-\gamma}}}{S^2(1-\gamma)^2}\left(\frac{S}{S_1}\right)^{\frac{1}{1-\gamma}}dS^2-2\frac{(1+\gamma) q^{-\frac{2}{1-\gamma}}}{S(1-\gamma)^2}\left(\frac{S}{S_1}\right)^{\frac{1}{1-\gamma}}dSdq\nonumber\\
&&-\frac{2(1+\gamma)}{(1-\gamma)^2}q^{\frac{3-\gamma}{1-\gamma}}\left(\frac{S}{S_1}\right)^{\frac{1}{1-\gamma}}dq^{2}\label{mw}\;.
\end{eqnarray}
The curvature scalar of this metric is identically null and this shows that the system can not be analysed as having or not phase transition. It is well known that this method is often inconsistent with the real physical thermodynamic system. In the next subsection we study thermodynamics through the method of Ruppeiner.

%%%%%%%%%%%%%%%%%%%%%%%%%%%%%%%

\subsection{ The Ruppeiner method}\label{subsec3.4}
%%%%%%%%%%%%%%%%%%%%%%%%%%%%%%%
Ruppeiner \cite{ruppeiner} also introduced a metric for defining the geometry of the thermodynamic space of the equilibrium states, but in this case, with the entropy as a thermodynamic potential. Again, this metric provides a curvature scalar $R_R$, which shows if the system possesses thermodynamic interaction and  phase transitions. The Ruppeiner metric is given by
\begin{eqnarray}
dl^{2}_{R(S)}&=&-\frac{\partial^2 S}{\partial M^2}dM^2-2\frac{\partial^2 S}{\partial M\partial q}dMdq-\frac{\partial^2 S}{\partial q^2}dq^2\nonumber\\
&=&S_{1}^{2} \gamma (1-\gamma)M^{-1-\gamma}q^{1+\gamma}dM^2-2S_1 (1+\gamma)(1-\gamma)M^{-\gamma}q^{\gamma}dMdq\nonumber\\
&&-S_{1}\gamma (1+\gamma)M^{1-\gamma}q^{-1+\gamma}dq^{2}\label{mr}\;.
\end{eqnarray}
The curvature scalar of this metric is identically null. This means that there is no thermodynamic interaction, which does not agree with the result obtained by the specific heat. It is also known that this method is often inconsistent with the thermodynamic description of the physical system.
\par
In the next subsection we will make the analysis of the thermodynamic system  by the  Liu-Lu-Luo-Shao method.

%%%%%%%%%%%%%%%%%%%%%%%%%%%%%%%

\subsection{ The Liu-Lu-Luo-Shao method}\label{subsec3.5}
%%%%%%%%%%%%%%%%%%%%%%%%%%%%%%%
\hspace{0,5cm}
Recently, Liu-Luo-Lu-Shao \cite{liu}, in the same order as Weinhold, has formulated a metric of the thermodynamic space of the equilibrium states, which provides a curvature scalar that determines whether the system possesses phase transitions. This metric is a new thermodynamic metric based on the Hessian matrix of several free energy, which in the case of the Helmholtz free energy, is given by
\begin{eqnarray}
dl^{2}_{LLLS(F)}&=&-dTdS+dA_{0}dq=-\frac{\partial T}{\partial S}dS^2+\left(\frac{\partial A_{0}}{\partial S}-\frac{\partial T}{\partial q}\right)dSdq+\frac{\partial A_{0}}{\partial q}dq^2\nonumber\\
&=&-\frac{\gamma T_{1}}{S(1-\gamma)}q^{-\frac{(1+\gamma)}{(1-\gamma)}}\left(\frac{S}{S_1}\right)^{\frac{\gamma}{1-\gamma}}dS^2-\frac{q^{\frac{-2}{1-\gamma}}}{S(1-\gamma)}\times\nonumber\\
&&\times\left(\frac{S}{S_1}\right)^{\frac{1}{1-\gamma}}\left[\bar{A}_{0}-(1+\gamma)T_1 S\right]dSdq\nonumber\\
&&+\frac{2\bar{A}_{0}}{(1-\gamma)}q^{-\frac{(3-\gamma)}{(1-\gamma)}}\left(\frac{S}{S_1}\right)^{\frac{1}{1-\gamma}}dq^2\label{mllls}\;.
\end{eqnarray}
The curvature scalar of this metric is identically null. As we had seen, this result also is not in accordance with that of the specific heat. An inconsistency with respect to the specific heat was shown for the topological case of black hole in Horava-Lifshitz gravity \cite{cao}. 
\par
We then see that for this class of solutions, all the more usual geometric methods that use a metric in the thermodynamic space of the equilibrium states, were incompatible with the results of the analysis made by the specific heat of the black holes. In the next section we will study the thermodynamic stability of this class of black hole solutions.

%%%%%%%%%%%%%%%%%%%%%%%%%%%%%%%

\subsection{ The local and global stability}\label{subsec3.6}
%%%%%%%%%%%%%%%%%%%%%%%%%%%%%%%
We now have to study the stability of this thermodynamic system. With respect to this, we may split into two parts, the local stability and the global one.
\par
The local stability is easily analysed by the specific heat (\ref{cqm})
\begin{equation}
C_{q}=\frac{(1-\gamma)}{\gamma}S\label{cqm1}\;.
\end{equation} 
Here, we see that for the normal case of EMD ($\eta_1=\eta_2=1$), the system is always locally stable for $0<\gamma<1$, i.e. $C_q>0$, and locally unstable for $-1<\gamma<0$. But for the phantom case, the system is always locally unstable, since $\gamma\in (-\infty, -1)$.
\par
Now we can analyse the global stability of the thermodynamic system. First, we define the Gibbs potential in the grand canonical ensemble 
\begin{eqnarray}
G=M-TS-\eta_2 A_0 q=M(1-T_1 S_1+\eta_2 \bar{A}_0)=\frac{M}{1-\gamma}\label{gibbs1}\;.
\end{eqnarray}
So, we see that there is no change of the Gibbs potential, unless for the variation of the parameter $\gamma$. For the normal case, $\gamma\in(-1,1)$, the potential is always positive and then, the system is globally unstable. But the phantom case, $\gamma\in (-\infty,-1)$, appears as the opposite of the previous one, and the system is globally stable. This is not consistent with our previous analysis, so that the Gibbs potential, for the grand canonical ensemble, does not seem suitable to do this analysis. 

On the other hand, the analysis made by the Helmholtz free energy in the canonical ensemble, shows a good agreement with the results already obtained in this work. We define the Helmholtz free energy as
\begin{eqnarray}
F=M-TS=M(1-T_1 S_1)=-\frac{\gamma M}{1-\gamma}\label{Hel}\;.
\end{eqnarray}
Now the situation seems to have been reversed. In the normal case, we have $F<0$ for $0<\gamma<1$, which leads to a globally stable system, such as the local stability made by the specific heat. However, the phantom case is always globally unstable.
\par
To conclude this section, we discuss the results obtained by specific heat and the Helmholtz free energy. 
\par
We see that there is no phase transition when we consider a specific solution with a fixed value of $\gamma$, for whatever the values of the extensive and intensive variables. But there is a transition of the solutions locally and globally stable to the unstable ones. This is particularly observed by observing the solution where $\gamma=0$, which separates the class of solutions locally and globally stable from that which are unstable. The specific heat (\ref{cqm1}) diverges in this case and the Helmholtz free energy (\ref{Hel}) is identically null. This can be seen directly by the metric (\ref{6}) that, for $\gamma=0$ yields the temperature 
$T=1/(4\sqrt{2}\pi q)$, the entropy  $S=4\sqrt{2}\pi Mq$ and the electric potential  $A_{0}=-M/q$. This shows that  $M=S/(4\sqrt{2}\pi q)=ST$, which, substituting in  (\ref{cqm}), leads to a divergence and in (\ref{Hel}) yields zero. Then, as  $C_q>0$ and $F<0$ for $0<\gamma<1$, $C_q\rightarrow \infty$ and $F=0$ for $\gamma=0$ and  $C_q<0$ and  $F>0$ for $\gamma\in(-\infty,-1)$, we get three phases with respect to the stability. For  $0<\gamma<1$ the thermodynamic of this part of class of locally and globally stable solutions, for $\gamma=0$ the system behaves as a point of phase transition, and for  $\gamma\in(-\infty,-1)$, the system is always locally and globally unstable.       
\par
Here we see a great similarity between our analysis and generalized solutions, for example in \cite{sheykhi6,sheykhi7}. In these papers, as well as in some other, we have stable and unstable phases, depending mainly on  the parameter of the coupling between the dilaton and the matter field, which in our case is that of Maxwell field. There exists a maximum value and one minimum to determine the stability of the system. For the parameter $\lambda$ in (\ref{action1}) (normal case $\eta_1=1$), we have a thermodynamic system globally and locally stable when $-1<\lambda<1$ ($0<\gamma<1$). For other values ​​of $\lambda$ or for the phantom case, we have an unstable system. So here we have no a Hawking-Page phase transition, as in \cite{sheykhi6,sheykhi7}.

%%%%%%%%%%%%%%%%%%%%%%%%%%%%%%%%%%%%%%%%%%%%%%%%%%%%%%%%%%%%%%%%%%%%%%%%%%%%%%%%%%%%%%%%%%%%%%
\section{ Conclusion}\label{sec4}

%%%%%%%%%%%%%%%%%%%%%%%%%%%%%%%%%%%%%%%%%%%%%%%%%%%%%%%%%%%%%%%%%%%%%%%%%%%%%%%%%%%%%%%%%%%%%%%%%%%%%%%%%%%%%%%%%%%%%%%%

\hspace{0,5cm} We re-obtained a class of non asymptotically flat black hole solutions of the EMD theory in 4D with spherical symmetry (\ref{6}). We also established a real interval for the parameter $\gamma$, such that $-1<\gamma<1$ is for the normal case and $\gamma\in(-\infty,-1)$ for the phantom case. We calculated the extensive and intensives variable for this thermodynamic system. The first law of thermodynamics is also established (\ref{pl}), by fixing the gauge of the electric potential, and calculated the specific heat (\ref{cqm}), or equivalently (\ref{cqs}).
\par
We analysed the thermodynamic system by the geometric methods of the geometrothermodynamics, Weinhold, Ruppeiner and that of  Liu-Lu-Luo-Shao. All analysis by the geometric methods provided an identically null curvature scalar for the thermodynamic space of the equilibrium states. This is incompatible with the analysis made by the usual specific heat, which provides an interacting system without extreme case nor phase transition.
\par
The local stability analysis done by the specific heat and the global stability made by the Helmholtz free energy in the canonical ensemble, shows that the normal case is locally and globally stable for $-1<\lambda<1$ ($0<\gamma<1$). On the other hand, for the other intervals of $\gamma$, including all phantom cases, the system proves unstable. The solution where $\gamma=0$ separates the class of locally and globally stable solutions from the unstable ones, but this can not be a Hawking-Page phase transition. Recently
Myung \cite{Myung} has found similar results to ours here, in the case of the Lanczos-Lovelock theory in d-dimensional AdS spacetimes. 
\par
We observe that the methods of geometrical analysis of the thermodynamic system proved incompatible with the specific heat due to the specific form of  the potentials and the intensive variables are written in terms of extensive variables. With this, our perspective is to change the metrics of the thermodynamic space of equilibrium states of these methods, in order to fit them with the particularity of this thermodynamic system, as had already been suggested for the geometrothermodynamics case in \cite{zui,deborah}.

%%%%%%%%%%%%%%%%%%%%%%%%%%%%%%%%%%%%%%%%%%%%%%%%%%%%%%%%%%%%

\vspace{0,5cm}

{\bf Acknowledgement}: M. E. Rodrigues thanks the UFES, UFRA and UFPA for the hospitality during the development of this work, thanks M. J. S. Houndjo for the help in the organization of the manuscript and  thanks CNPq for partial financial support. G.T.M. thanks CNPq and  FAPESPA for partial financial support.

%%%%%%%%%%%%%%%%%%%%%%%%%%%%%%%%%%%%%%%%%%%%%%%%%%%%%%%%%%%%%%%%%%%%%%%%%%%%%%%%%%%%%%%%%%%%%%%%%%%%%%%%%%%%%%%%%%%%%%%%%%%%%%%%%

%%%%%%%%%%%%%%%%%%%%%%%%%%%%%%%%%%%%%%%%%%%%%%%%%%%%%%%%%%%%%%%%%%%%%%%%%%%%%%%%%%%%%%%%%%%%%%
\end{document}